\documentclass[twocolumn,showpacs,aps,prd,superscriptaddress,preprintnumbers,amsmath,amssymb,floatfix]{revtex4}
\usepackage{graphicx}

\newcommand{\BABARPubYear}    {05}
\newcommand{\BABARPubNumber}  {42}
\newcommand{\SLACPubNumber} {11431}

\usepackage{relsize}
\usepackage{xspace}
\def\babar{\mbox{\slshape B\kern-0.1em{\smaller A}\kern-0.1em
    B\kern-0.1em{\smaller A\kern-0.2em R}}}
\def\pep2{PEP-II}
    
\def\g     {\ensuremath{\gamma}\xspace}
\def\gaga  {\ensuremath{\gamma\gamma}\xspace}  
\def\piz   {\ensuremath{\pi^0}\xspace}
\def\Kp    {\ensuremath{K^+}\xspace}
\def\Km    {\ensuremath{K^-}\xspace}
\def\pip   {\ensuremath{\pi^+}\xspace}
\def\epem    {\ensuremath{e^+e^-}\xspace}
\def\Dstarz  {\ensuremath{D^{*0}}\xspace}
\def\Dz      {\ensuremath{D^0}\xspace}
\def\Ds      {\ensuremath{D^+_s}\xspace}
\def\Dss     {\ensuremath{D^{*+}_s}\xspace}

\def\invfb   {\ensuremath{\mbox{\,fb}^{-1}}\xspace}
\def\Y#1S{\ensuremath{\Upsilon{(#1S)}}\xspace}
\def\FourS {\Y4S}
\def\BR         {{\ensuremath{\cal B}\xspace}}
\def\lsim{{~\raise.15em\hbox{$<$}\kern-.85em
          \lower.35em\hbox{$\sim$}~}\xspace}

\newcommand{\stat}{\ensuremath{\mathrm{(stat.)}}\xspace}
\newcommand{\syst}{\ensuremath{\mathrm{(syst.)}}\xspace}

\newcommand{\mev}{\ensuremath{\mathrm{\,Me\kern -0.1em V}}\xspace}
\newcommand{\mevc}{\ensuremath{{\mathrm{\,Me\kern -0.1em V\!/}c}}\xspace}
\newcommand{\mevcc}{\ensuremath{{\mathrm{\,Me\kern -0.1em V\!/}c^2}}\xspace}
\newcommand{\gevc}{\ensuremath{{\mathrm{\,Ge\kern -0.1em V\!/}c}}\xspace}

\newcommand{\jprlBase}       {Phys.\ Rev.\ Lett.\xspace}
\newcommand{\jprl}      [1]  {\jprlBase\ {\bf #1}}
\newcommand{\jprBase}        {Phys.\ Rev.\xspace}
\newcommand{\jprd}      [1]  {\jprBase\ D~{\bf #1}}
\newcommand{\jplBase}        {Phys.\ Lett.\xspace}
\newcommand{\plb}       [1]  {\jplBase\ B~{\bf #1}}
\newcommand{\nimBaseA}       {Nucl.\ Instr.\ Meth.\xspace}
\newcommand{\nima}      [1]  {\nimBaseA~A~{\bf #1}}

\newcommand{\ppmm}[2]{\penalty900\mkern5mu^{+\penalty900\mkern5mu{#1}}_{-\penalty900\mkern5mu{#2}}}

\begin{document}

\preprint{BABAR-PUB-\BABARPubYear/\BABARPubNumber}
\preprint{SLAC-PUB-\SLACPubNumber}

\title{ \boldmath
Measurement of the Branching Ratios 
$\Gamma(\Dss\to\Ds\piz)/\Gamma(\Dss\to\Ds\g)$
and $\Gamma(\Dstarz\to\Dz\piz)/\Gamma(\Dstarz\to\Dz\g)$
}

%
\author{B.~Aubert}
\author{R.~Barate}
\author{D.~Boutigny}
\author{F.~Couderc}
\author{Y.~Karyotakis}
\author{J.~P.~Lees}
\author{V.~Poireau}
\author{V.~Tisserand}
\author{A.~Zghiche}
\affiliation{Laboratoire de Physique des Particules, F-74941 Annecy-le-Vieux, France }
\author{E.~Grauges}
\affiliation{IFAE, Universitat Autonoma de Barcelona, E-08193 Bellaterra, Barcelona, Spain }
% ...
\collaboration{The \babar\ Collaboration}
\noaffiliation

\date{\today}

\begin{abstract}
Data samples corresponding to 
the isospin-violating decay $\Dss\to\Ds\piz$ and the
decays $\Dss\to\Ds\g$, $\Dstarz\to\Dz\piz$
and $\Dstarz\to\Dz\g$ are 
reconstructed using $90.4$~$\invfb$ of data recorded by the \babar\  
detector at the \pep2 asymmetric-energy $e^+e^-$ collider. The
following branching ratios are extracted:
$\Gamma(\Dss\to\Ds\piz)/\Gamma(\Dss\to\Ds\g) = 0.062\pm0.005\:\stat\pm0.006\:\syst$
and
$\Gamma(\Dstarz\to\Dz\piz)/\Gamma(\Dstarz\to\Dz\g) = 1.74\pm0.02\:\stat\pm0.13\:\syst$.
Both measurements represent significant
improvements over present world averages.
\end{abstract}

\pacs{13.25.Ft, 13.40.Hq, 12.39.Fe}

\maketitle

The decay of any higher-mass
$c\bar{s}$ meson into  $\Ds\piz$~\cite{footnote:ChargeConj}
violates isospin conservation, thus guaranteeing a small partial width. 
The amount of suppression is a matter of large theoretical uncertainty
according to most models of charm-meson radiative decay~\cite{bib:GoityRoberts}.
One such model~\cite{bib:ChoWise} suggests that the decay 
$\Dss\to\Ds\piz$ may proceed via 
$\piz$-$\eta$ mixing. Even including such considerations,
the radiative decay $\Dss\to\Ds\g$ is still expected to dominate. 
The existence of
isospin-violating decay modes such as $\Dss\to\Ds\piz$
is particularly relevant given the recent
observations of two narrow new \Ds\ meson states~\cite{bib:Ds232,bib:Ds246}.
In particular, in contrast to the $\Dss$ meson,
there is no experimental evidence for the electromagnetic
decay of the  $D_{sJ}(2317)^+$; current measurements place the branching
ratio to $\Dss\gamma$
at less than 18\% at 90\% confidence level (CL)~\cite{bib:Ds246belle}.

Besides the $\Ds\piz$ and $\Ds\gamma$ final states,
no other decay modes of the $\Dss$ have been observed
and none are expected to occur at a significant level. 
Only one previous observation of the decay $\Dss\to\Ds\piz$  is
recorded in the literature, yielding a value of
$0.062\ppmm{0.020}{0.018}\:\stat\pm0.022\:\syst$ for the
branching ratio $\Gamma(\Dss\to\Ds\piz)/\Gamma(\Dss\to\Ds\g)$~\cite{bib:CLEO}.
The analysis presented here confirms this observation
and provides a more precise measurement of this branching ratio.

The decay $\Dstarz\to\Dz\piz$, in contrast to $\Dss\to\Ds\piz$, 
does not violate isospin conservation and the world average for the 
branching ratio is
$\Gamma(\Dstarz\to\Dz\piz)/\Gamma(\Dstarz\to\Dz\g)=1.625\pm0.20$~\cite{bib:PDG2004}.
As for the $\Dss$ meson, the $\piz$ and $\gamma$ decay modes are expected to 
saturate the decay width of the $\Dstarz$ meson.

The results presented here are based on data recorded by the 
\babar\ detector at the \pep2 asymmetric-energy \epem\ storage rings.
The data sample, corresponding to an integrated luminosity of
$90.4\invfb$, was 
recorded at and approximately $40$~\mev\  below the $\FourS$ resonance.
Due to the unequal beam energies, the $e^+e^-$ center-of-mass system
is boosted relative to the laboratory frame with $\beta\gamma \approx 0.55$.
The \babar\ detector and trigger 
are described in detail elsewhere~\cite{bib:NIM}.
Charged particles are detected and their momenta measured by a
silicon vertex tracker (SVT) consisting of five layers of double-sided
silicon strip sensors and a cylindrical 40-layer drift chamber (DCH), both
operating within a $1.5\mathrm{\,T}$ solenoidal magnetic field. 
Charged particle identification is provided by energy loss measurements
in the SVT and DCH and by Cherenkov light
detected in an internally reflecting ring 
imaging detector (DIRC).
Photons are identified and their energies measured
by an electromagnetic
calorimeter (EMC) composed of 6580 CsI(Tl) crystals.

In the following paragraphs,
the $\Dss$ measurement is described first. The
$\Dstarz$ analysis, which uses similar procedures for signal extraction,
is described afterward in less detail.

\Ds\ mesons are reconstructed via the decay sequence
$\Ds\to\phi\pip$, $\phi\to\Kp\Km$.
Kaons are identified by combining the energy deposited in
the SVT and DCH with the information from the DIRC.
Tracks not
identified as kaons according to the particle identification 
criteria are considered to be pions.
All $\Kp\Km\pip$ candidates are required to fit successfully to a common
vertex. Only combinations with a $\Kp\Km$ invariant
mass within $8$~\mevcc\  of the
nominal $\phi$ mass~\cite{bib:PDG2004} are retained.

In \epem\ annihilation to charm quarks, the $c\bar{c}$ fragmentation
process is characterized by the production of high-momentum (leading)
charm hadrons. This property is exploited in order to reduce substantially
the combinatorial background by
retaining only those $\phi\pip$ candidates with scaled momentum
$x_p$ greater than $0.6$, where $x_p$ is defined as $x_{p}(\Ds) = p^{*}(\Ds) /
p^{*}_{\mathrm{max}}(\Ds)$ and $p^{*}(\Ds)$ is the momentum of the \Ds\
candidates in the $e^+e^- $center-of-mass frame 
with $p^{*}_{\mathrm{max}}(\Ds) =
\sqrt{{E^{*}_{\mathrm{beam}}}^{2} - {m(\Ds)}^{2}}$ as its maximum value.

The longitudinal polarization of the $\phi$ meson 
in the \Ds\  rest frame is used
to reduce background by requiring
that the absolute value of the cosine of the helicity angle,
defined as the angle between the $\phi$ momentum direction in the
\Ds\ rest frame and the momentum direction of either of the kaons
in the $\phi$ rest frame, is $0.3$ or greater.

\begin{figure}
\includegraphics[width=\columnwidth]{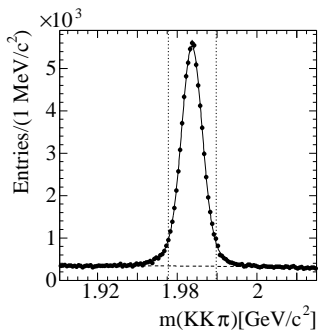}
\caption{\label{fig:Ds}The $\Kp\Km\pip$ mass distribution.
The dots represent data points with error bars corresponding to
statistical uncertainties (these uncertainties 
are small enough that the error bars
are difficult to distinguish). 
The solid curve shows the fitted function.
The dashed curve indicates the 
background. \Ds\ candidates are defined by the region between the
vertical dotted lines.}
\end{figure}

The resulting $K^+K^-\pip$ invariant mass distribution is 
shown in Fig.~\ref{fig:Ds}.
This distribution can be modeled by the 
sum of two Gaussian functions (to represent the signal) and
a third-order polynomial (to represent the background). The resulting
binned $\chi^2$ fit yields $73\,500\pm300$ events (statistical errors only). 
A \Ds\ candidate is retained if its
invariant mass is within $12$~\mevcc\  of the nominal 
\Ds\ mass~\cite{bib:PDG2004}.

A \piz\ candidate is reconstructed by combining two photon candidates
that fulfill the following requirements.
Each photon candidate is identified by a
calorimeter cluster that is not associated with a charged track and has an
energy in the laboratory frame of at least $45$~\mev.
Additionally, to help remove the background from hadronic showers, 
the fractional lateral width~\cite{bib:LAT}, which describes the shape of the
shower in the calorimeter, is required to be less than $0.55$.
The fiducial acceptance of photon candidates is restricted by
the angular range of the EMC ($-0.92 \lsim \cos\theta \lsim 0.89$, 
where $\theta$
is the polar angle in the center-of-mass frame~\cite{bib:NIM}).

A \piz\ candidate is retained if it has a momentum $p^{*}$ in the
$e^+e^-$ center-of-mass frame greater 
than $150$~\mevc. Furthermore, the absolute
value of the cosine of the decay angle, $\theta^{*}$, which is
defined as the angle between the direction of one of the photons in the
\piz\ rest frame and the direction of the \piz\ candidate in the
center-of-mass frame, is required to be less than $0.85$. For 
$\piz\to\gamma\gamma$ decay, the $\cos \theta^{*}$ distribution is uniform,
while it peaks near $\pm 1$ for random $\gaga$ combinations.

Only $\gaga$ pairs within a specified mass interval are retained. This 
interval is
defined by the values of mass at which the $\piz$ signal portion of
a function fitted to the $\gaga$ mass
distribution falls below 0.2 times its
maximum value. This requirement 
accommodates the asymmetric shape of the $\gaga$ mass
distribution and takes into account variations in detector calibration. A
kinematic fit is applied to the surviving $\gaga$ pairs to constrain
their mass to the nominal \piz\ mass.

\begin{figure}
\includegraphics[width=\columnwidth]{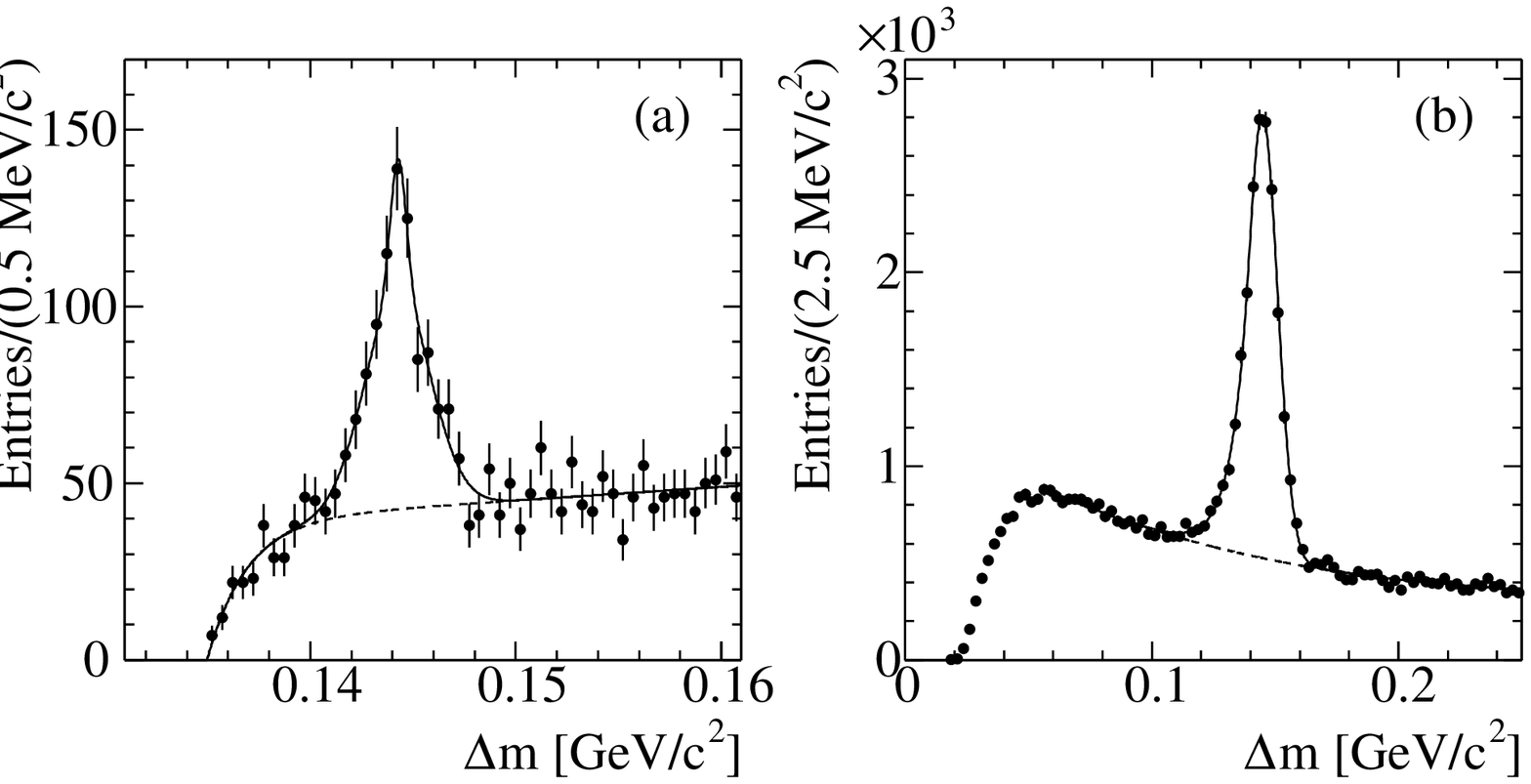}
\caption{\label{fig:Dss}The \Dss\ signals:
(a)~$m(\Kp\Km\pip\piz)-m(\Kp\Km\pip)$;
(b)~$m(\Kp\Km\pip\g)-m(\Kp\Km\pip)$.
The dots represent data points. The solid curve shows the fitted function.
The dashed curve indicates the portion of the fit associated with
background.}
\end{figure}

After combining the \Ds\ and \piz\ candidates in a search for the decay
$\Dss\to\Ds\piz$, a fit is applied to the distribution of the mass difference
$\Delta m (\Ds\piz) = m(\Kp\Km\pip\piz) - m(\Kp\Km\pip)$. The fit function
is the sum of a double Gaussian function to represent the signal and
the function
\begin{eqnarray}
f_{1}(\Delta m) &=& N \left( 1 - \exp \left( - \frac{\Delta m - m(\piz)}{\mu} \right) \right) \nonumber \\
                &\times& \left( \Delta m^{2} + a\Delta m + b \right) \;,
\label{eq:ExpQuad}
\end{eqnarray}
where $m(\piz)$ is the \piz\ mass, and $N$, $\mu$, $a$, and $b$ are free
fit parameters to describe the background. The exponential term 
models the kinematic threshold; this threshold
term has little influence on the background shape near the signal region.
The result of this fit is shown in
Fig.~\ref{fig:Dss}(a). A signal event yield of $560\pm40$
(statistical error only) is obtained.

For the reconstruction of the decay $\Dss\to\Ds\g$, a calorimeter cluster that
is not associated with a charged track is considered a photon candidate if it
fulfills the following requirements: the energy must be $50$~\mev\  or greater in
the laboratory frame and $100$~\mev\  or greater in the 
$e^+e^-$ center-of-mass frame and
the fractional lateral width 
must be less than $0.8$. To reduce the background due to
photons from \piz\ decay, a photon candidate is discarded if it forms a \piz\
candidate with any other photon candidate in the same event. In this
case, a $\gaga$
combination is considered a \piz\ candidate if the invariant mass is in the
range $115 < m(\gaga) < 155$~\mevcc\  and if
the total energy is at least $200$~\mev\ in the 
$e^+e^-$ center-of-mass frame.

To obtain the $\Dss\to\Ds\g$ signal event yield, a fit is applied
to the distribution of 
the mass difference $\Delta m (\Ds\g) = m(\Kp\Km\pip\g) - m(\Kp\Km\pip)$.
The fit function is a sum of a third-order polynomial to model the background
plus a function first introduced by the Crystal Ball 
collaboration~\cite{bib:CrystalBall} for the signal
\begin{equation}
f_{2}(\Delta m) = N \cdot \left\{
  \begin{array}{ll}
  A \left( B - \frac{\Delta m-\mu}{\sigma} \right) ^{-n}         & {\mathrm{if}}\:(\Delta m-\mu)/\sigma \le \alpha \\
  \exp \left( - \frac {(\Delta m-\mu)^{2}}{2 \sigma^{2}} \right) & {\mathrm{if}}\:(\Delta m-\mu)/\sigma \ge \alpha
  \end{array} \right.\;,
\label{eq:CrystalBall}
\end{equation}
where $N$, $\mu$, $\sigma$, $n$, and $\alpha$ are free fit parameters
and $A$ and $B$ are chosen such that the function and its first derivative
are continuous at $(\Delta m-\mu)/\sigma = \alpha$. 
The fit result is shown
in Fig.~\ref{fig:Dss}(b). A signal yield of $15\,600\pm200$ events
(statistical error only) is obtained.

The reconstruction efficiencies are 
determined using a Monte Carlo simulation based
on $30\,000$ events for each $\Ds$ decay mode. The simulated events 
are analyzed using the same procedure as for real data. By
calculating the ratio of the number of reconstructed to generated events,
efficiencies of $\epsilon(\Ds\piz) = 0.041\pm0.002$
and $\epsilon(\Ds\g) = 0.071\pm0.002$ are found for the two \Dss\ decay
modes. The efficiency ratio is
$\epsilon(\Ds\piz)/\epsilon(\Ds\g) = 0.58\pm0.03$
(statistical error only).

Various sources of systematic uncertainties are studied.
To verify that the Monte Carlo events model the data correctly, 
$\tau$ decays with one or two \piz\ mesons in the final state are
studied to
obtain energy-dependent Monte Carlo efficiency corrections for
\piz\ mesons and photons. Although this procedure indicates that
no correction is necessary, the errors on
the correction functions represent uncertainties in the Monte Carlo model
and contribute a systematic uncertainty of 3.6\%.

To test for uncertainties in the background shape of the mass difference
distributions, upper and lower sidebands in the $\Kp\Km\pip$ and 
$\gamma\gamma$ mass distributions 
are considered. Positive signal yields are expected
in these sidebands from either mis-reconstructed or
unassociated $\piz$ candidates. To measure these yields, the same fit
functions used to determine the signal yields are applied 
to the mass difference
distributions of the sideband samples. Any discrepancy in yield
so obtained from data and Monte Carlo simulation is considered a systematic
uncertainty (4.8\%). Most of this uncertainty is attributed to 
the relatively large background in the $\Ds\piz$ decay mode.

\begin{figure}
\includegraphics[width=\columnwidth]{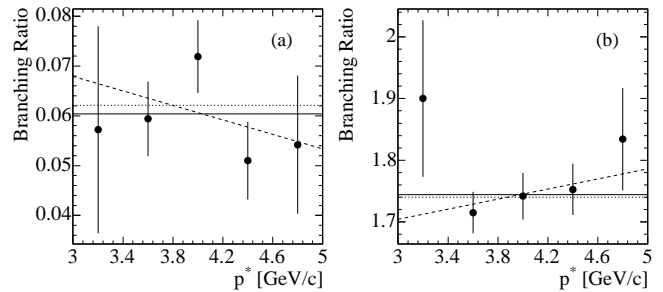}
\caption{\label{fig:ratio}The measured values of
(a)~$\Gamma(\Dss\to\Ds\piz)/\Gamma(\Dss\to\Ds\g)$ and
(b)~$\Gamma(\Dstarz\to\Dz\piz)/\Gamma(\Dstarz\to\Dz\g)$
in intervals of center-of-mass momentum
$p^*$. The error bars indicate the 
associated statistical error.
The solid line (dashed line) is
the result of a fit to a constant (first-order polynomial).
The dotted line is the result from the entire sample integrated over $p^*$.}
\end{figure}

The measurement of $\Gamma(\Dss\to\Ds\piz)/\Gamma(\Dss\to\Ds\g)$
is repeated
for the subsamples of candidates within various $p^*$ intervals. 
By fitting either a constant function
or a first-order polynomial to the branching ratio as a
function of $p^{*}$, it is possible to verify
that the measured branching ratios
are independent of $p^{*}$ (see Fig.~\ref{fig:ratio}).
Nevertheless, it is assumed conservatively that
any $p^*$ dependence arises from unknown momentum dependencies of the
efficiencies that may not cancel in the branching ratios.
The difference (6.8\%)
between the branching ratio represented by
the constant function and the integral of the first-order polynomial 
is therefore reported as a systematic
uncertainty. 

The systematic uncertainties are summarized in Table~\ref{tb:systs}.
Combining all contributions in quadrature,
a total systematic uncertainty of 10.2\% is derived
for the measurement of $\Gamma(\Dss\to\Ds\piz)/\Gamma(\Dss\to\Ds\g)$.

\begin{table}
\caption{\label{tb:systs}A summary of the relative
systematic uncertainties in the branching ratio measurements.}
\begin{ruledtabular}
\renewcommand{\baselinestretch}{1.3}
\begin{tabular}{lrrrr}
        & \multicolumn{4}{c}{Relative Uncertainty (\%)} \\
\cline{2-5}
 & 
\multicolumn{2}{c}{$\Gamma(\Dss\to\Ds\piz)$}  & 
\multicolumn{2}{c}{$\Gamma(\Dstarz\to\Dz\piz)$}  \\
\cline{2-3}
\cline{4-5}
Sources & 
\multicolumn{2}{c}{$\Gamma(\Dss\to\Ds\gamma)$}  & 
\multicolumn{2}{c}{$\Gamma(\Dstarz\to\Dz\gamma)$}  \\
\hline
Background shape         & \hspace{24pt} 4.8 && \hspace{24pt} 0.1 &\\
Monte Carlo statistics   &  5.0 &&  5.4 &\\
Signal model             &  3.6 &&  3.8 &\\
$p^*$ dependence         &  6.8 &&  2.8 &\\ 
\hline
Quadrature Sum           & 10.2 &&  7.2 &\\
\end{tabular}
\end{ruledtabular}
\end{table}

The ratio
$\Gamma(\Dstarz\to\Dz\piz)/\Gamma(\Dstarz\to\Dz\gamma)$,
where $\Dz\to\Km\pip$, is measured
using the same
selection criteria for the \piz\ and photon candidates
as in the reconstruction of $\Dss\to\Ds\piz$ and $\Dss\to\Ds\g$. 
To be included in the $\Dz\to\Km\pip$ sample, 
a candidate \Km\ and \pip\ combination must yield an acceptable
fit to a common vertex and
the scaled momentum $x_p$ of the resulting \Dz\ candidate must be $0.6$ or
greater. Fitting the sum of a double Gaussian function and a third-order
polynomial to the resulting $\Km\pip$ invariant mass distribution
(not shown) produces
$(996.0\pm1.5)\times10^{3}$ signal events (statistical error only). 
A $\Km\pip$ combination is
retained if its mass differs by less than $17$~\mevcc\  from the nominal
\Dz\ mass~\cite{bib:PDG2004}.

\begin{figure}
\includegraphics[width=\columnwidth]{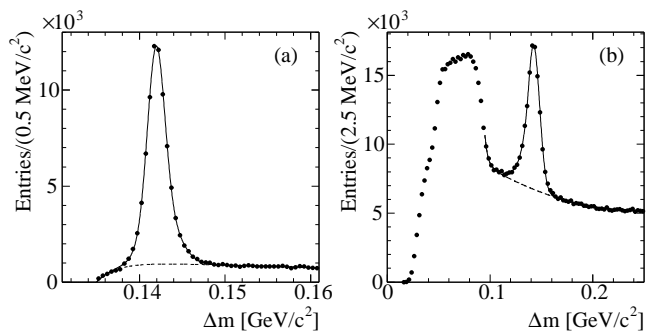}
\caption{\label{fig:Dstarz}The \Dstarz\ signals:
(a)~$m(\Km\pip\piz)-m(\Km\pip)$ and
(b)~$m(\Km\pip\g)-m(\Km\pip)$.
The dots represent data points. The solid curve shows the fitted function.
The dashed curve indicates the portion of the fit associated with
background.}
\end{figure}

The \Dz\ candidates are combined with all \piz\ candidates; the resulting
mass difference $\Delta m (\Dz\piz) = m(\Km\pip\piz) - m(\Km\pip)$ is
shown in Fig.~\ref{fig:Dstarz}(a). A fit using a double Gaussian for the
signal and the function shown in Eq.~\ref{eq:ExpQuad} for the background
yields $69\,000\pm450$ signal events (statistical error only).

The \Dz\ candidates are then combined with all photon candidates producing
the distribution of the mass difference $\Delta m (\Dz\g) = m(\Km\pip\g) - m(\Km\pip)$
shown in Fig.~\ref{fig:Dstarz}(b). In this case, the  peak corresponding to
$\Dstarz\to\Dz\g$ signal is close to a large bump arising
from the reflection of 
$\Dstarz\to\Dz\piz$
in which one photon is produced by \piz\ decay (the same reflection appears
in $\Dss$ decay but with a lower rate and less distinctive shape). 
Most of this bump is avoided by limiting the analysis to
$\Delta m > 95$~\mevcc. The remainder of
the background is modeled using the function
\begin{eqnarray}
f_{3}(\Delta m) &=& N \left( 1 + \exp \left( - \frac{\Delta m - m(\piz)}{\mu} \right) \right) \nonumber \\
                &\times& \left( \Delta m^{2} + a\Delta m + b \right) .
\end{eqnarray}
(Note that this function is similar to that of Eq.~\ref{eq:ExpQuad},
but differs in the sign of the exponential term.)
The signal is modeled by the Crystal Ball function (Eq.~\ref{eq:CrystalBall}).
The resulting fitted signal consists of $67\,880\pm670$ events
(statistical errors only).

Efficiencies and systematic uncertainties  are determined using the
procedures described 
for $\Gamma(\Dss\to\Ds\piz)/\Gamma(\Dss\to\Ds\g)$.  Efficiencies
of $\epsilon(\Dz\piz) = 0.037\pm0.002$ and $\epsilon(\Dz\g) = 0.064\pm0.002$
and an efficiency ratio of $\epsilon(\Dz\piz)/\epsilon(\Dz\g) = 0.58\pm0.03$ 
are found. The latter is consistent with the value of
$\epsilon(\Ds\piz)/\epsilon(\Ds\g)$. The ratio 
$\Gamma(\Dstarz\to\Dz\piz)/\Gamma(\Dstarz\to\Dz\g) = 1.74\pm0.02\:\stat\pm0.13\:\syst$ 
is obtained.

The branching ratio measurements are summarized in Table~\ref{tab:Results}.
By assuming that the \Dss\ meson decays only to $\Ds\piz$ and $\Ds\g$,
and that the
\Dstarz\ meson decays only to $\Dz\piz$ and $\Dz\g$,
it is possible to calculate the branching fractions, which are also listed
in Table~\ref{tab:Results}.

In summary, the branching ratio
$\Gamma(\Dss\to\Ds\piz)/\penalty-900\Gamma(\Dss\to\Ds\g) = 0.062\pm0.005\:\stat\pm0.006\:\syst$
has been measured and 
is consistent with the previous measurement~\cite{bib:CLEO}, but has higher
precision. Also determined is the ratio
$\Gamma(\Dstarz\to\Dz\piz)/\Gamma(\Dstarz\to\Dz\g) = 1.74\pm0.02\:\stat\pm0.13\:\syst$.
This result is in agreement with, but is more precise than,
the world average~\cite{bib:PDG2004}.

It has been proposed that the decay $\Dss\to\Ds\piz$
proceeds via $\eta-\piz$ mixing 
and calculations based on
Chiral perturbation theory~\cite{bib:ChoWise} predict 
$\BR(\Dss\to\Ds\piz) \approx 1$--$3\%$ based on current measurements
of $\BR(D^{*+}\to D^+\gamma) = 1.6 \pm 0.4\%$~\cite{bib:PDG2004}.
Newer theoretical estimates in a relativistic quark 
model~\cite{bib:GoityRoberts} 
predict $\BR(\Dss\to\Ds\piz) \approx 13\%$,
somewhat larger than our measurement.

\begin{table}
\caption{\label{tab:Results}Summary of the results. 
The first errors are statistical;
the second represent systematic uncertainties.}
\begin{ruledtabular}
\begin{center}
\begin{tabular}{ll@{$\:\pm\:$}l@{$\:\pm\:$}l}
$\Gamma(\Dss\to\Ds\piz)/\Gamma(\Dss\to\Ds\g)$       & $0.062$& $0.005$ &$0.006$ \\
$\BR(\Dss\to\Ds\piz)$                               & $0.059$& $0.004$ &$0.006$ \\
$\BR(\Dss\to\Ds\g)$                                 & $0.942$& $0.004$ &$0.006$ \\ \hline
$\Gamma(\Dstarz\to\Dz\piz)/\Gamma(\Dstarz\to\Dz\g)$ & $1.74  $& $0.02  $ &$0.13 $  \\
$\BR(\Dstarz\to\Dz\piz)$                            & $0.635 $& $0.003 $ &$0.017$ \\
$\BR(\Dstarz\to\Dz\g)$                              & $0.365 $& $0.003 $ &$0.017$ \\
\end{tabular}
\end{center}
\end{ruledtabular}
\end{table}

   We are grateful for the 
extraordinary contributions of our \pep2\ colleagues in
achieving the excellent luminosity and machine conditions
that have made this work possible.
The success of this project also relies critically on the 
expertise and dedication of the computing organizations that 
support \babar.
The collaborating institutions wish to thank 
SLAC for its support and the kind hospitality extended to them. 
This work is supported by the
US Department of Energy
and National Science Foundation, the
Natural Sciences and Engineering Research Council (Canada),
Institute of High Energy Physics (China), the
Commissariat \`a l'Energie Atomique and
Institut National de Physique Nucl\'eaire et de Physique des Particules
(France), the
Bundesministerium f\"ur Bildung und Forschung and
Deutsche Forschungsgemeinschaft
(Germany), the
Istituto Nazionale di Fisica Nucleare (Italy),
the Foundation for Fundamental Research on Matter (The Netherlands),
the Research Council of Norway, the
Ministry of Science and Technology of the Russian Federation, and the
Particle Physics and Astronomy Research Council (United Kingdom). 
Individuals have received support from 
CONACyT (Mexico),
the A. P. Sloan Foundation, 
the Research Corporation,
and the Alexander von Humboldt Foundation.


\begin{thebibliography}{9}
\bibitem{footnote:ChargeConj}
The use of charge conjugate states is implied throughout this paper.
\end{thebibliography}
\end{document}